\documentclass[10pt, notitlepage]{revtex4-1}
\usepackage[ruled, vlined, linesnumbered]{algorithm2e}
\usepackage{circuitikz}
\usepackage{tikz}
\usetikzlibrary{calc, intersections, through, backgrounds, arrows, positioning, automata, shapes, shapes.geometric, fit}
\usepackage{amsmath}
\usepackage{amssymb}
\usepackage{geometry} 
\usepackage{array}
\usepackage{booktabs}
\usepackage{graphicx}
\usepackage{caption}
\usepackage{listings}
\usepackage{fancyhdr}
\usepackage{yfonts}
\usepackage{mathrsfs}
\usepackage[mathscr]{euscript}
\usepackage{relsize}
\usepackage{bigints}
\usepackage{verbatim} 
\usepackage[english]{babel}
\usepackage{slantsc}
\usepackage{amsthm}
\usepackage{bbm}
\usepackage{placeins}
\usepackage{cancel}
\usepackage{setspace}
\usepackage{mathtools}
\usepackage{blkarray}
\usepackage{multirow}
\usepackage{bigdelim}
\usepackage{xcolor}

\DeclareMathAlphabet{\mathpzc}{OT1}{pzc}{m}{it}

\usepackage{calligra}
\usepackage{rotunda}
\usepackage{mflogo}
\usepackage{frcursive}
\usepackage{suetterl}

\input Sanremo.fd

\input Rothdn.fd

\newcommand{\mfr}[1]{\text{\fontfamily{frc}\selectfont\bfseries\footnotesize#1}\,}

\newcommand{\msuet}[1]{\text{\Large\suetterlin#1}}

\newfont{\rsfsten}{rsfs10 scaled 1200}

\newcommand{\n}{\vspace{\baselineskip}}

\newcommand{\sm}{\smallskip} 
\newcommand{\m}{\medskip} 

\newcommand{\gt}{\rightarrow} 

\newtheoremstyle{propositional}
{10pt}
{10pt}
{
	\addtolength{\linewidth}{-2.0em}
	\parshape 1 2.0em \linewidth} 
{}
{\sc}
{:}
{.5em}
{}
\makeatother

\theoremstyle{propositional}

\newtheoremstyle{definitive}
{10pt}
{10pt}
{
	\addtolength{\linewidth}{-2.0em}
	\parshape 1 0.0em \linewidth} 
{}
{\sc}
{:}
{.5em}
{}
\makeatother

\theoremstyle{definitive}

\newcounter{numb}
\newcounter{bean}
\newcounter{count}
\newcounter{count2}
\newcounter{count3}

\makeatletter

\newlength\tdima
\newcommand\tabfill[1]{%
	\setlength\tdima{\linewidth}%
	\addtolength\tdima{\@totalleftmargin}%
	\addtolength\tdima{-\dimen\@curtab}%
	\parbox[t]{\tdima}{#1\ifhmode\strut\fi}}

\newcommand\mytabs{\hspace*{5cm}\=\hspace{1cm}\=\hspace{2cm}}

\makeatother

\definecolor{gainsboro}{RGB}{220,220,220}
\definecolor{textclr}{RGB}{131,148,150}
\definecolor{monokai}{HTML}{272822}
\definecolor{darkb}{RGB}{0, 43, 54}

\tikzset{elliptic state/.style={draw,ellipse}}

\DontPrintSemicolon

\begin{document}

\title{Quantum Annealing with Special Drivers for Circuit Fault Diagnostics}

\date{\today}

\author{Hannes Leipold$^{1,2,\ast}$, Federico M. Spedalieri$^{1,3}$}

\affiliation{
	$^1$Information Sciences Institute, University of Southern California, Marina del Rey, CA 90292, USA \\
	$^2$Department of Computer Science, University of Southern California, Los Angeles, CA 90089, USA \\
	$^3$Department of Electrical and Computer Engineering, University of Southern California, Los Angeles, CA 90089, USA \\
	$\ast$Corresponding author: Hannes Leipold, leipold@usc.edu
}

\begin{abstract}
We present a very general construction for quantum annealing protocols to solve Combinational Circuit Fault Diagnosis (CCFD) problems that restricts the evolution to the space of valid diagnoses.  This is accomplished by using special local drivers that induce a transition graph on the space of feasible configurations that is regular and instance independent for each given circuit topology. Analysis of small instances shows that the energy gap has a generic form, and that the minimum gap occurs in the last third of the evolution. We used these features to construct an improved annealing schedule and benchmarked its performance through closed system simulations. We found that degeneracy can help the performance of quantum annealing, especially for instances with a higher number of faults in their minimum fault diagnosis. This contrasts with the performance of classical approaches based on brute force search that are used in industry for large scale circuits.

\end{abstract}

\maketitle

\section{Introduction}

\quad Quantum computation can offer new and exciting prospects for solving problems that 
are difficult with existing classical approaches~\cite{harrow2017quantum}\cite{arute2019quantum}. 
Quantum Annealing (QA) and the Quantum Approximate Optimization Algorithm~\cite{farhi2014quantum} are two popular 
approaches for doing quantum computation on near-term quantum devices~\cite{gyongyosi2019quantum}\cite{gyongyosi2019survey}\cite{alexeev2021quantum}\cite{preskill2021quantum}.
While Driver Hamiltonians in QA and ansatz strategies in QAOA are often selected based on their 
simplicity to be implemented on near-term devices, an exciting prospect to improve the performance
of these systems is to tailor these constructions to the underlying structure of the problem being solved~\cite{bian2016mapping}\cite{cerezo2021variational}\cite{hadfield2021representation}. Many modern quantum computational frameworks attempt to leverage quantum phenomena to solve combinatorial optimization problems. Among these is quantum annealing (QA), which leverages adiabatic (or diabatic) evolution from one Hamiltonian to another, usually through a standard interpolation. Applications in \textit{combinational circuit fault diagnostics} (CCFD) have gained considerable attention~\cite{perdomo2015quantum}\cite{bian2016mapping}\cite{perdomo2019readiness} for quantum annealers to potentially leverage quantum speed up. Developments in Constrained Quantum Annealing (CQA)~\cite{hen2016quantum}\cite{hen2016driver}\cite{leipold2021constructing} and Quantum Alternating Operator Ansatz (QAOA)~\cite{hadfield2017quantum}\cite{hadfield2019quantum}\cite{wang2020x}\cite{shaydulin2020classical}\cite{hadfield2021analytical} have centered around more novel driver operators, with higher body interaction terms than the transverse field. As noted in Ref.~\cite{bian2016mapping}, driver Hamiltonians beyond the transverse field are pertinent to understanding what QA can bring to solving such problems and this work can be seen as a step in this direction. 

\m 

\quad In this paper we delineate a CQA approach for solving single input/output CCFD problems that maintains support \textit{only} on the space of \textit{valid diagnoses}. By constraining the anneal to only explore \textit{feasible} states associated with a given CCFD instance, we find a square root reduction in the size of the explored space compared to the standard transverse field approach in the explicit mapping setting~\cite{perdomo2019readiness}. It also avoids the need to impose energy penalties on the final Hamiltonian to suppress invalid diagnoses, that in turn results in stringent requirements for the range of the interaction parameters. The driver terms that are needed to do this are all local (with bounded weight), representing the possible  transformations that takes a feasible configuration to another feasible configuration involving only the variables associated with a single circuit gate. These driver terms are sufficient and minimal for describing a driver Hamiltonian that constrains the evolution to the subspace of all valid diagnoses, such that they form a \textit{generating set} for all possible driver Hamiltonians that constrain the evolution to this feasibility space~\cite{leipold2021constructing}. 

\m

\quad To solve an instance of the CCFD problem we need, (i) an initial 1-local Hamiltonian that has as its ground state the trivial diagnosis (all faults occur in the output gates), (ii) a driver Hamiltonian composed of the driver terms discussed in the previous paragraph, and (iii) a final Hamiltonian (also 1-local) that favors diagnoses with the smaller number of faults. Utilizing a standard interpolation between these three Hamiltonians, the wavefunction of the system begins with support on a specific valid configuration and then explores all valid configurations while the evolution of the instantaneous Hamiltonian to the final Hamiltonian eventually restricts the wavefunction to support on the subset of diagnoses with the minimum number of faults. 

\m 

\quad The paper is organized as follows. In Section~\ref{sec:cfd}, we provide background on the combinatorial problem CCFD and in Section~\ref{sec:qa} delineate how our approach can be utilized to solve this problem. In Sec.~\ref{sec:benchmarks}, we analyze the minimum gap of the annealing spectrum for a set of randomly generated non-degenerate instances of a modified version of the C17 circuit form the ISCAS benchmarks~\cite{brglez1985neural}, as well as the success probability for different choices of the annealing schedule. We then construct a larger, richer family of instances on which we run a generic but improved annealing schedule, and study the relationship between the success probability, the degeneracy of the ground space of the instance, and the number of faults of the minimum fault diagnosis. While these instances require 52 qubits to represent explicitly, because the closed system evolution occurs in a much smaller space, we are still able to tractably simulate these systems on a workstation. The results suggest our approach can be beneficial for employing quantum annealers to solve instances where higher number of faults are needed to be able to diagnose the circuit. 

\section{Combinatorial Optimization for Circuit Fault Diagnostics}\label{sec:cfd}
\begin{center}
	\begin{figure}
		\includegraphics[width=0.500\textwidth]{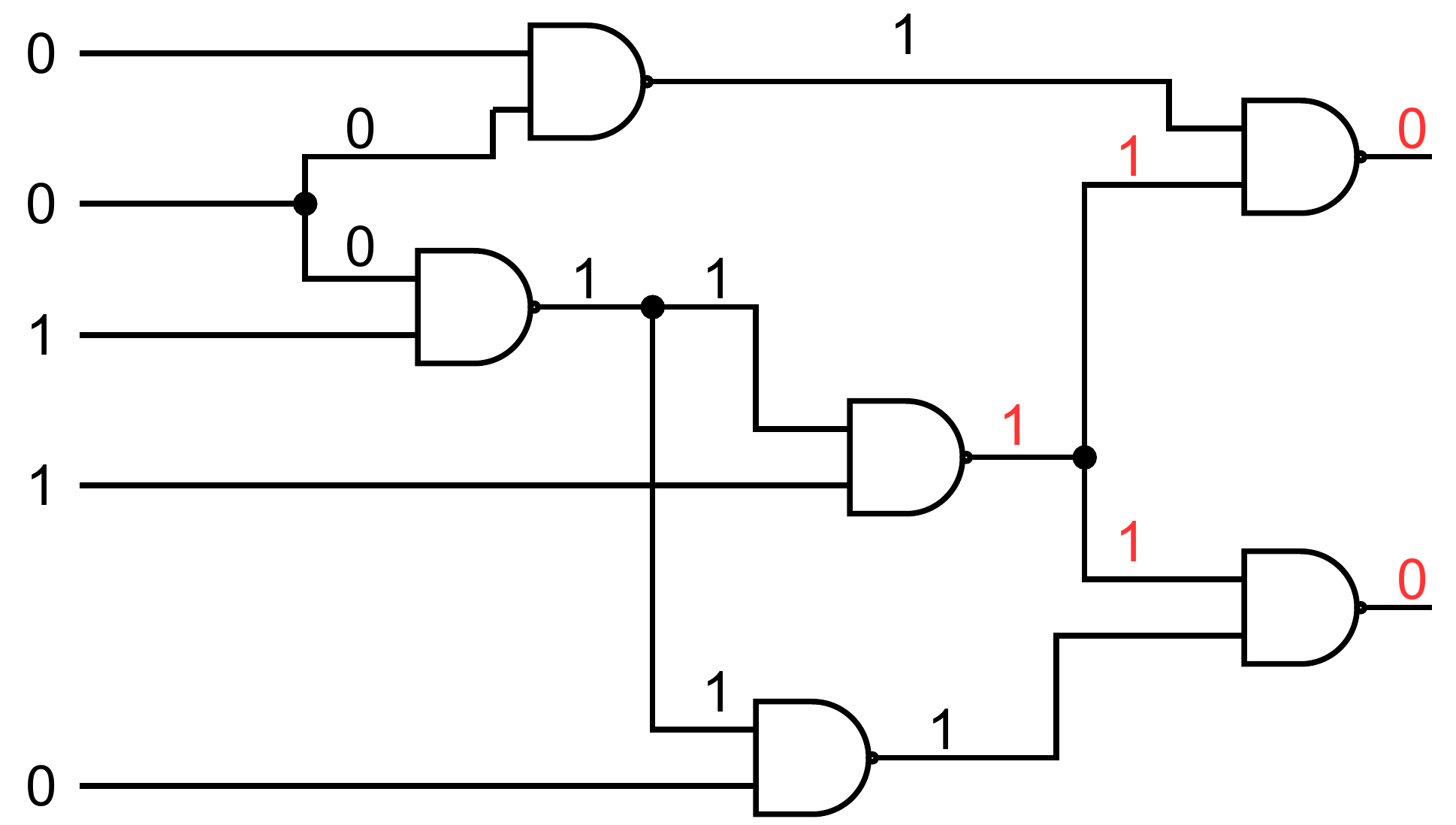}
		\caption{A schematic of the C17 circuit with a fault on output of the third NAND gate for a given input/output pair. In Sec.~\ref{sec:benchmarks}, we generate new random circuits by replacing each NAND gate with a randomly selected two input logic gate.}\label{fig:c17}
	\end{figure}
\end{center}
\vspace{-2.5em}

\quad Consider a combinational circuit with $N$ wires (or locations), with $N_I$ inputs and $N_O$ outputs. Each wire can take one of two values, $x_i = 0$ or $x_i = 1$, so a state of the circuit is associated with an $N$ bit string or assignment, $(x_1,\ldots,x_N)$. The wires are connected to logical gates in such a way that any wire that is not an input or an output for the entire circuit, is connected to exactly two gates (it is the output of one gate and the input to another one). Inputs (outputs) to the circuit are inputs (outputs) to (for) one gate. The connectivity of the gates determines the function performed by the circuit. This is referred to as the structural description of the circuit, and is one of the most commonly used abstractions in testing digital systems \cite{gupta2002}.

\m 

\quad For the purpose of our discussion here, we consider gates that are either fan out gates (FAN), which maps a single input to \textit{two} outputs such that the two outputs have values equivalent to the input, an inverter gate (INV) or one of the following logic gates: OR, AND, XOR, NOR, NAND. The methods described in this paper can be extended trivially to a more general $ k_{1}-$inputs, $ k_{2}-$outputs gate.  

\m 

\quad Since we are using the structural description of the circuit, we will use the most well known structural fault model, the stuck-at fault (SAF) model of CCFD \cite{gupta2002}. In this model, a wire is either healthy, or fails by being either stuck-at-0 (SA0) or stuck-at-1 (SA1). Given an input-output pair, and a description of the circuit structure, a diagnosis is an explanation of this input-output relationship that is consistent with the chosen fault model. To describe a diagnosis, we will add to the $N$ bits describing the values of each wire, another $N$ fault bits $f_i$, associated with each of the wires, that would indicate if a wire is faulty ($ f_i = 1 $) or not ($ f_i = 0 $). 

\m 

\quad Under the assumption that every wire is equally likely to be faulty, the goal of the circuit fault diagnosis (CFD) problem is to find the minimal number of faults necessary to provide a consistent explanation for the outputs of the circuit~\cite{perdomo2015quantum}. We will refer to any diagnosis with the minimum number of faults as a \textit{minimum fault diagnosis} (MFD). Note that our representation does not specify the type of fault (SA0 or SA1), but this information can be inferred from the state of the wire: if a wire has a value $ x_i = 1 $ ($ x_i = 0 $) and $ f_i =1 $, we can conclude the wire must be SA1 (SA0). 

\subsection{The set of valid diagnoses}

\quad We can refer to any $2N-$bit string as a diagnosis, in the sense that it provides a description of the state of the wires on the system and the location of potential faults. However, not any $2N-$bit string will represent a fault assignment that is consistent with the chosen fault model. We will call this subset of consistent assignments, the set of  \textit{valid diagnoses}.  Hence, a valid diagnosis can be represented by a $2N-$bit string, $(x_{1}, \ldots, x_{n}, f_{1}, \ldots, f_{n})$, where the first $N$ bits describe the state of the circuit, while the last $N$ bits flag whether a wire is healthy or not. For every wire location $ i $, if the assignment value $ x_{i} $ is not compatible with the logic of the gate that it is an output to and the corresponding inputs for that gate, then the fault flag $ f_{i} $ must be set to 1 for the assignment to valid, otherwise the assignment is invalid (as a false negative). Likewise, if the assignment value $ x_{i} $ is compatible, then the fault flag $ f_{i} $ must be set to 0, otherwise the assignment is invalid (as a false positive). 

\m 

\quad The last $N_O$ wires correspond to the output bits of the circuit and are fixed by the problem instance. Given a collection of empirical outputs for a faulty circuit, there are potentially many valid diagnoses to explain this phenomenon, since the faults could occur anywhere in the circuit, but a fault occurring on a particular wire does not mean the subsequent wire values are faulty since they were compatible with their respective gates (i.e. the inputs for the gate were different from what the non-faulty circuit would have and this led to their erroneous assignment). The first $N_I$ wires correspond to the inputs and are in principle fixed. However, we must allow for a fault to occur before the input wires reach the first gates, so the first $ N_I $ wires are allowed to take any value: any discrepancy with the actual inputs to the circuit will be reflected by the corresponding fault bit being set to 1. The consistency of the fault assignment is imposed on the wires and fault bits associated with every gate in the circuit. We can illustrate this with the following example.

\m     

\quad Consider the bit string $(x_{i_1}, x_{i_2}, x_{o}; f_{i_1}, f_{i_2}, f_{o}) $ representing the wires and fault bits associated with a NAND gate in the circuit. The configuration $(110;000)$ represents a consistent assignment, in the sense that the output of the gate is compatible with its inputs, i.e., $ x_o = \text{NAND}(x_{i_1} , x_{i_2}) $, and no wire is flagged as faulty; $ (100;001) $ is also a consistent assignment, because even though the output wire is inconsistent with the inputs and the gate function ($ x_o \neq \text{NAND}(x_{i_1} , x_{i_2}) )$, the output wire is flagged as faulty ($f_o = 1$). Examples of inconsistent assignments would be $ (111;000) $ and $ (011;001) $, where in the first case the output is incorrect but its fault bit indicates no fault, while in the second case the output is correct but the fault bit is flagging it as faulty. Note that the fault bit of a gate input is of no importance in deciding whether the output of that gate is faulty or not: the output is faulty (healthy) if it is inconsistent (consistent) with the values of the inputs of the gate that produces it. A valid diagnosis will then be described by a $2N-$ bit string, where for all circuit gates, the corresponding wires and fault bits form a consistent assignment for the specific function of that gate.
In Table \ref{tab:nandvalids} we present an exhaustive list of all consistent assignments for a NAND gate, indexed by the values of the input and output wires. 
\begin{center}
    \begin{table}
		\begin{tabular}{ | c | c | c | }
			\hline  &    \\
		    ($x_{i_1}, x_{i_2}; x_{o}$) & Consistent assignments ($x_{i_1}, x_{i_2}, x_{o}; f_{i_1}, f_{i_2}, f_{o}$)\\
			 & \\
			\hline
			 & \\
			 (0,0;0)   &   (0,0,0;0,0,1), (0,0,0;1,0,1), (0,0,0;0,1,1), (0,0,0;1,1,1)    \\
			 (1,0;0)   &   (1,0,0;0,0,1), (1,0,0;1,0,1), (1,0,0;0,1,1), (1,0,0;1,1,1)    \\
			 (0,1;0)   &   (0,1,0;0,0,1), (0,1,0;1,0,1), (0,1,0;0,1,1), (0,1,0;1,1,1)  \\
			 (1,1;0)   &   (1,1,0;0,0,0), (1,1,0;1,0,0), (1,1,0;0,1,0), (1,1,0;1,1,0)   \\
			 & \\
			\hline
		 & \\ 
		   (0,0;1)   &   (0,0,1;0,0,0), (0,0,1;1,0,0), (0,0,1;0,1,0), (0,0,1;1,1,0)   \\
		   (1,0;1)   &   (1,0,1;0,0,0), (1,0,1;1,0,0), (1,0,1;0,1,0), (1,0,1;1,1,0)   \\
		   (0,1;1)   &   (0,1,1;0,0,0), (0,1,1;1,0,0), (0,1,1;0,1,0), (0,1,1;1,1,0)  \\
		   (1,1;1)  &    (1,1,1;0,0,1), (1,1,1;1,0,1), (1,1,1;0,1,1), (1,1,1;1,1,1)  \\
		 & \\ 
			\hline
		\end{tabular}
		\caption{Consistent assignments for a NAND gate: note that there is no restriction on the values of the inputs, their fault bits and the output (hence, we have $2^5 = 32$ configurations). However, the output fault bit is completely determined by the values of the inputs and the output. }
		\label{tab:nandvalids}
	\end{table}
\end{center}
\m 

\quad We can now see how to construct any possible valid diagnosis. Given any arbitrary value for the first $N-N_{O}$  wires, and the values of the $N_O$ output wires, we can then univocally set the value of each fault bit $f_i$ such that it is consistent with the value of the wire and the constraint imposed by the preceding gate. Hence the set of valid diagnoses has $2^{N-N_O}$ elements.

\section{Constrained Quantum Annealing Approach}\label{sec:qa}

\quad Quantum Annealing (QA) has been used in many combinatorial optimization tasks~\cite{farhi2000quantum}\cite{hen2011exponential}\cite{rieffel2015case}\cite{king2019quantum}. Many studied quantum annealing protocols involve two Hamiltonians: a driver Hamiltonian like the transverse field $ H_{d} = - \sum_{i} \sigma_{i}^{x} $ and a final Hamiltonian made of single body and two body spin-z interaction terms $ H_{f} = \sum_{i} \left( h_{i} \sigma_{i}^{z} + \sum_{j} J_{ij} \sigma_{i}^{z} \sigma_{j}^{z} \right) $~\cite{perdomo2019readiness} ($\sigma_{i}^{x}$ and $\sigma_{i}^{z}$ are the standard Pauli matrices). A typical linear transverse field annealing schedule will then be $ H(s(t)) = (1 - s(t)) H_{d} + s(t) \, H_{f} $ with the time function $ s(t) = t / T_{f} $ for some specified total time $ T_{f} $. The transverse field driver Hamiltonian mixes configurations that differ by a single bit flip. Applying it to our CFD problem would result in generating configurations that are not valid diagnoses. One way to deal with this issue is to add a term to the final Hamiltonian that penalizes any configuration that is not a valid diagnosis. This is a commonly used technique to deal with constraints both in classical and quantum optimization~\cite{lucas2014ising}\cite{venturelli2015quantum}. In QA this introduces other problems, such as a determining the right energy scale for these penalty terms, and requiring more qubit connectivity to represent them. Constrained Quantum Annealing (CQA)~\cite{hen2016quantum}\cite{hen2016driver}\cite{leipold2021constructing} and the Quantum Alternating Operator Ansatz (QAOA)~\cite{hadfield2017quantum}\cite{hadfield2019quantum}\cite{wang2020x}\cite{shaydulin2020classical}\cite{hadfield2021analytical} have been introduced to avoid these precise issues, and our work aligns with the spirit of these approaches.

\m 

\quad  In the case of CQA, a ground state for the driver Hamiltonian can be more difficult to prepare and so annealing protocols are often generalized beyond a two Hamiltonian interpolation. To overcome this barrier, we consider annealing protocols involving three Hamiltonians, a form for annealing schedules often seen when exploiting `catalysts'~\cite{albash2019role}\cite{albash2021diagonal}\cite{cao2021speedup}: $ H(s(t)) = A(s(t)) \, H_{i} + B(s(t)) \, H_{d} + C(s(t)) \, H_{f} $ with $ s:[0,T_{f}] \gt [0,1] $, $ A(s(0)) = 1 - B(s(0)) = 1 - C(s(0)) = 1 $, and $ A(s(T_{f})) = B(s(T_{f})) = 1 - C(s(T_{f})) = 0 $. In Sec.~\ref{subsec:specialdrivers} we show how to explicitly construct the driver Hamiltonian for this approach and in Sec.~\ref{subsec:implement_cqa} describe a general CQA protocol using an interpolation of an initial Hamiltonian, the driver Hamiltonian, and a final Hamiltonian.

\subsection{Special Drivers for CQA Solvers tackling CFD}\label{subsec:specialdrivers}

\quad The special CFD drivers are designed to locally connect pairs of  valid diagnoses. To each gate in the circuit we will associate operators that are non-diagonal in the computational basis, which couple two valid diagnoses that differ only on the qubits associated with that specific gate. Since each input and output wire of a gate is mapped to two qubits (one for the wire value and one for the fault bit), the special drivers will have a weight that is double the number of inputs plus outputs of the gate. 

\m 

\quad We will now show how to construct the required drivers for a NAND gate (this process can be extended to any other gate). Assume that we have a valid diagnosis for the whole circuit, and let $| x_{i_1} x_{i_2} x_o ; f_{i_1} f_{i_2} f_o \rangle$ be the corresponding state restricted to the qubits associated with this specific NAND gate. We want to find other configurations of these 6 qubits that also represent a valid diagnosis. The transformed configuration must be locally and globally consistent with the assignment of other wires and fault bits in the circuit. To ensure global consistency, we must (i) keep the value of the output wire fixed, to prevent propagating its effect forward, and (ii) flip the wire and associated fault bit of any input \textit{simultaneously}, since this will keep the new assignment consistent with any previous gate for which this input wire is an output. Summarizing, we can only change the input wires and their fault bits (as long as the wire and the corresponding fault bit are flipped together), and the output fault bit to ensure the local consistency of the new diagnosis. 

\m 

\quad  To illustrate the idea, consider a valid diagnosis whose state restricted to the qubits representing a given NAND gate are given by the state $|100;001\rangle$, where the output wire is flagged as faulty because it does not respect the gate's functionality. If we flip $x_{i_1}$ and $f_{i_1}$, we get $|000;101\rangle$ and the output fault bit remains unchanged since this configuration still represents a failing NAND gate. Same if we flip both inputs and their fault bits. However, if we flip only $x_{i_2}$ and $f_{i_2}$, the resulting configuration $|110;011\rangle$ has an output wire that now is consistent with the NAND gate, but its fault bit marks it as faulty. Hence, we should also flip the output fault bit to $0$ to make it consistent, resulting in the new local configuration $|110;010\rangle$.  Gathering all this information, we see that the required driver terms to implement these local transformations of a valid diagnosis will be:
\begin{equation}
	\{|000;101\rangle \langle 100;001| + h.c., |110;010\rangle \langle 100;001|+ h.c. , |010;111\rangle\langle 100;001|+ h.c.\}, \label{eq:exsetdrivers}
\end{equation}
 where \textit{h.c.} stands for Hermitian conjugate. The same analysis can be carried out for any valid configuration of the NAND gate. There are 32 such valid configurations for a 2-input, 1-output gate (cfr. Table \ref{tab:nandvalids}). For each one of those we need 3 driver terms (per valid configuration) to perform the input flips (and flip the output bit if necessary). But due to the hermiticity requirement, the total number of driver terms is reduced by half, to 48.

\subsection{Implementing CQA protocols for CFD with special drivers}\label{subsec:implement_cqa}

\quad In order to map an instance of CFD to our quantum annealing framework, we need to specify an initial Hamiltonian, a driver Hamiltonian and a final Hamiltonian. The driver Hamiltonian can be constructed from the circuit description, as a linear combination of the drivers corresponding to each gate in the gate set $ G $ of the circuit. There is actually some freedom on what the coefficients of this linear combination could be (as long as they are nonzero)~\cite{leipold2021constructing}. This could potentially impact the performance of the approach and developing useful strategies for choosing these coefficients is an intriguing direction for future research. Here we will discuss this briefly later when we compare the effects of stoquastic versus non-stoquastic drivers, but for now we will proceed with the following (stoquastic) assignment
\begin{equation}
	H_{D} = -\sum_{ g \in G} H_{g},  
	\label{eq:stoqdriver}
\end{equation}
where each $H_g$ is a sum of the drivers associated with the gate $g$, which we described how to construct in the previous section (see Eq.~\ref{eq:exsetdrivers}).  Each gate has a bounded number of drivers associated with it, so the total number of terms in $ H_{D} $ scales linearly with the number of gates in the circuit. 

\m 

\quad The initial Hamiltonian should have a \textit{ground state}, written in the computational basis, that is a valid diagnosis. Such a state can always be found by simply assuming that all gates are healthy except for the gates that produce the output bits that are known to be wrong. The input and output wires are fixed by the instance, and the interior wire values are computed by propagating the input through the healthy circuit. The fault bits are set to 0, except for the ones corresponding to the faulty output wires, which are set to 1. 

\m 

\quad Let $|x_1\ldots x_N f_1 \ldots f_N \rangle$ be that initial state. Then we can initialize the system by implementing an initial Hamiltonian diagonal in the computational basis given by 
\begin{equation}
	H_{I} = \sum_{i=1}^{N} ((-1)^{x_i} \, \sigma_i^z  +   (-1)^{f_i} \, \sigma_{N+i}^z ),
\end{equation}
which has this initial state as its ground state. The final Hamiltonian must penalize diagnoses that have a higher number of faults, so the Hamiltonian
\begin{equation}
	H_{F} = -\sum_{i=N+1}^{2N} \sigma_i^z,
\end{equation}
where the sum is only over the fault bit qubits. We can then combine all this into a time dependent Hamiltonian:
\begin{equation}
	H(s(t)) = A(s(t))\, H_{I} + B(s(t))\, H_{D} + C(s(t))\, H_{F}. \label{eq:instham}
\end{equation}
Hence, the system starts in a specific valid diagnosis, and once the driver is turned on it will explore \textit{only the space of valid diagnoses}. Finally, at the end, the final Hamiltonian will favor diagnoses with the lowest number of faults. In comparison to the transverse field approach used with explicit mappings in Refs.~\cite{bian2016mapping}\cite{perdomo2015quantum}\cite{perdomo2019readiness}, there is no need for penalty terms associated with constraining the solution to be a valid diagnosis, which also requires setting a coefficient in the order of $ N_{O} $ for the most general type of instances. 

\m 

\quad The action of the driver Hamiltonian allows us to reach any valid diagnosis from any other valid diagnosis, so the adiabatic evolution generates a superposition of all valid diagnoses. The graph induced by $ H_{D} $ in this space is connected and regular, with degree $ (3 G_{2} + G_{1}) $, where $ G_{2} $ and $ G_{1} $ are the number of gates with 2 and 1 inputs, respectively. Furthermore, for a given connectivity, the corresponding induced graphs are isomorphic for all instances (i.e., any input-output pairs). This is due to the fact that the driver terms are essentially just performing all possible flips of the inputs of a gate, while the fault bits are flipped accordingly to enforce the constraints of a valid diagnoses. In the next Section, we benchmark the CQA approach discussed on small sized circuits.

\section{Benchmarks on Synthetic Instances: Spectral Gaps and Simulated Annealing Schedules}\label{sec:benchmarks}

\subsection{Generalized ISCAS C17 Random Circuits: Spectral Gaps and Parameterized Annealing Schedule}\label{subsec:c17}

\quad Our first example is a circuit known as C17 (see Fig.~\ref{fig:c17}), that appeared in the ISCAS'85 benchmarks~\cite{brglez1985neural}\cite{hansen1999unveiling}. This circuit has 17 wires, 5 inputs, 2 outputs, 6 NAND gates and 3 FAN gates. Using our approach, it requires 34 qubits to represent all the wire values and the fault bits. The reduced dynamics occurs on  a space of dimension $2^{15}$, so it is possible to simulate the closed system evolution and compute the spectrum along the anneal. 

\m 

\quad We generated new circuit instances by randomly replacing the NAND gates with other 2-input, 1 output gates, randomly chosen from the set \{NAND, AND, OR, NOR, XOR\}. Then we randomly chose the 5 input bits, simulated the circuit to compute the 2 output bits, and then flipped the value of these 2 outputs. The CFD instance is then defined by the circuit description, and the input-output pair. Note that the number of output bits that are faulty is always an upper bound on the number of faults in the \textit{minimal fault diagnosis} (MFD), since assigning the fault only to those output bits is itself a valid diagnosis. That is also the reason why we flipped both output bits, to make room for a non-trivial MFD with a single fault.

\subsubsection{Spectral Gaps for C17 Instances}
\begin{center}
\begin{figure}
	\begin{tabular}{ c c }
	\includegraphics[width=0.500\textwidth]{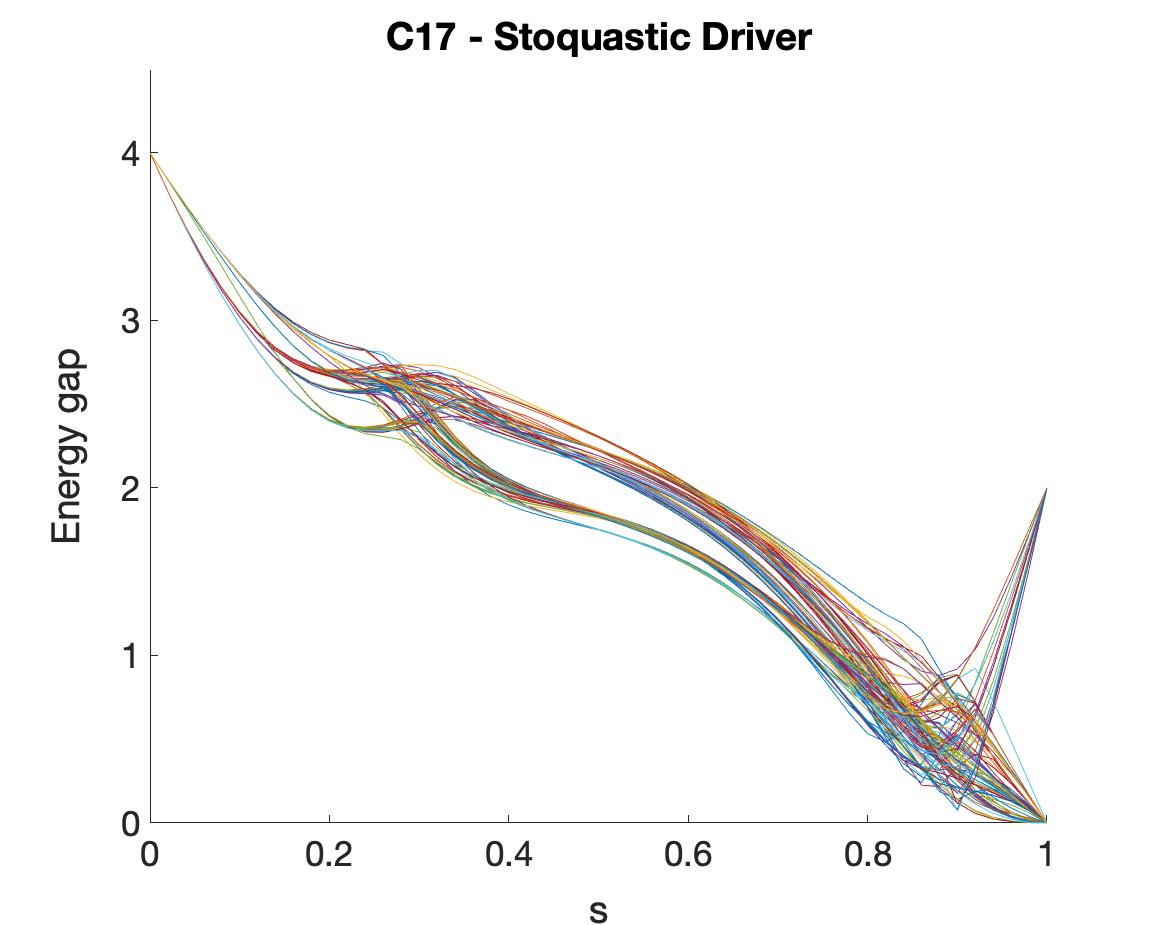} & \includegraphics[width=0.550\textwidth]{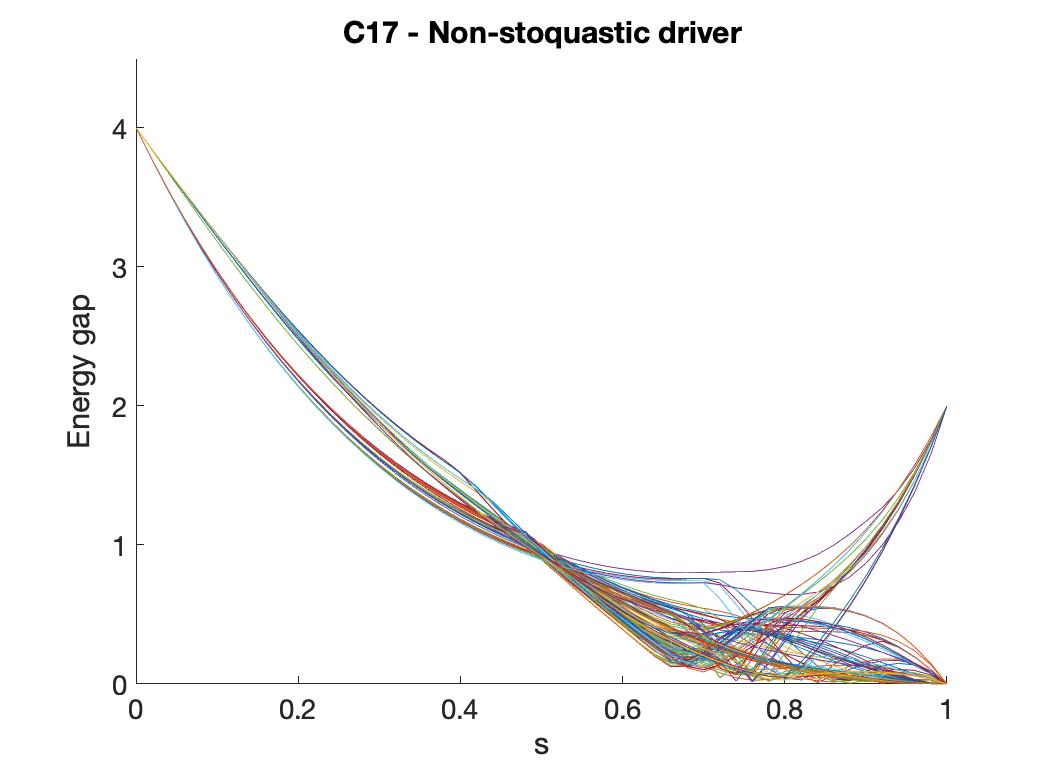} \\ 
	(1) & (2) 
	\end{tabular}
	\caption{(1) shows the instantaneous minimum gap for 100 random instances of the generalized C17 circuit at regular discretized points of time. (2) similarly shows the instantaneous ground state gap for the same instances with a non-stoquastic version of the driver Hamiltonian.}\label{fig:spectrum_stoq_n_nonstoq}
\end{figure}
\end{center}
\vspace{-2.5em}
\quad Fig.~\ref{fig:spectrum_stoq_n_nonstoq}.1 shows the energy gap between the ground state and the first excited state at every point during the anneal, for 100 randomly generated instances. We can see that the gap always starts at 4 (due to the form of the initial Hamiltonian, and the fact that any two valid diagnoses are at least Hamming distance 2 away). Depending on the degeneracy of the instance, the gap either has an absolute minimum during the anneal (non-degenerate), or goes to 0 at the end of the anneal (degenerate). An interesting feature is that the minimum gap for non-degenerate instances seems to be localized on the last third of the anneal. The generic shape of the gap as a function of the anneal parameter could be exploited to evolve the system faster at the beginning, when the gap is large. This would only result in a constant factor reduction in the total annealing time required, so it has no impact on the complexity, but is useful for simulating larger circuits.

\m 

\quad We also computed the spectrum corresponding to flipping the sign of the Hamiltonian driver, which produces a non-stoquastic Hamiltonian. Due to the infamous sign problem, non-stoquastic Hamiltonians are thought to be harder to simulate than stoquastic ones~\cite{albash2019role}\cite{crosson2016simulated}, leading some to conjecture they may provide some computational speedup in some situations and non-stoquastic catalyst Hamiltonians, such as considered in Ref.~\cite{albash2019role}, are able to achieve large improvements over catalyst-free approaches in the finite and infinite~\cite{nishimori2017exponential} range ferromagnetic p-spin model. Ref.~\cite{choi2021essentiality} likewise shows possible improvements from introducing non-stoquastic terms in a driver Hamiltonian. However, there is also evidence~\cite{crosson2020signing} that they can produce smaller energy gaps, resulting in a computational slowdown. 

\m 

\quad In Fig.~\ref{fig:spectrum_stoq_n_nonstoq}.2 we show the minimum gap for the non-stoquastic version of the driver Hamiltonian. Comparing this with Fig.~\ref{fig:spectrum_stoq_n_nonstoq}.1 we can see that in this case, the minimum gap seems to be generically smaller, and the region where the gap is small is also larger. It is important to point out that these are not the only choices for the coefficients of the  terms in the  driver Hamiltonian. Since these coefficients determine how different diagnoses are mixed by the driver, one could imagine that a better choice, perhaps one that exploits the particular structure of the circuit, may favor some sort of interference that could enhance diagnoses with lower number of faults and result in a computational advantage. However, it is not clear at this point if that is feasible, or even what the complexity of finding those coefficient would be. It is nonetheless an intriguing direction for future research.

\subsubsection{Choosing annealing schedule to exploit spectrum features}
\label{sec:ansched}
\begin{center}
\begin{figure}
	\makebox[\textwidth][c]{
	\begin{tabular}{ c c }
	\includegraphics[width=0.5\textwidth]{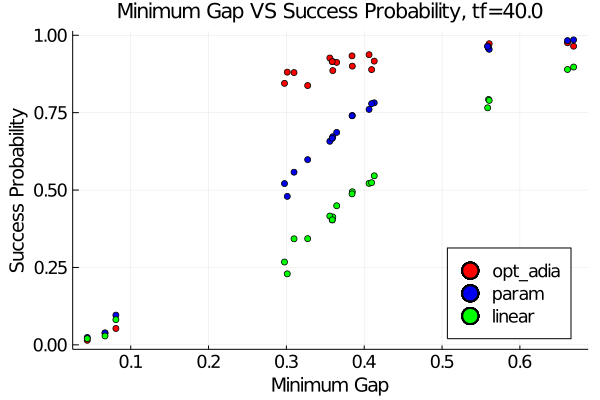} & \includegraphics[width=0.5\textwidth]{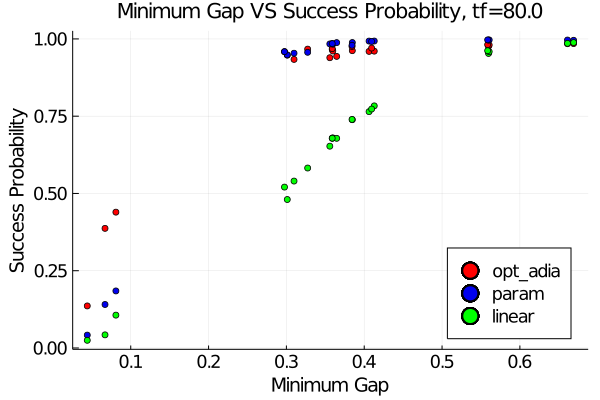} \\
	(1) & (2)
	\end{tabular} }
	\caption{(1) and (2) show the success probability versus the minimum gap for 22 non-degenerate single fault instances of the C17, with 40 units and 80 units of time respectively, for the single parameter function (labeled param), the linear function (labeled linear) and a piece-wise linear function fitted over 100 evenly spaced points, such that the slope is proportional to the instantaneous inverse gap squared between evenly spaced $ s_{i-1} $ and $ s_{i} $ (labeled opt\_adia).} \label{fig:lin_v_param_v_best}
\end{figure}
\end{center}
\vspace{-2.5em}

\quad For our annealing schedule, we chose functions $ A(s(t)) = 1 - s^{2}(t) $, $ B(s(t)) = 4 s \, (1-s) $, and $ C(s(t)) = s^{2}(t) $ (the instantaneous Hamiltonian is then given by Eq.~\ref{eq:instham}). 
We utilized the generic features of the energy gap for C17 (and in Sec.~\ref{subsec:26} for C26), to generate a single parameter annealing schedule, that evolves faster at the beginning of the anneal (when the gap is larger), and then slows down when traversing the region containing the minimum gap.  For a given amount of total anneal time $ T_f $, the annealing schedule assigns $ T_{0} $ ($T_0 < T_f$) units of time for the range $ [ 0, s_{0} ] $, following a quadratic function, and then shifts at point $ T_{0} $ to a linear function over the interval $ [ s_{0}, 1 ] $ such that $ s'(T_{0}) $ is not a point of discontinuity. These conditions yield:
\begin{align*}
    s(t) =  \begin{cases}
                \left( \frac{ s_{0} \left( T_{f} / T_{0} \right) - 1 }{ T_{0} \left( T_{0} - T_{f} \right) } \right) t^{2} + \left( \frac{1 - s_{0} \left( 2 \, \left( T_{f}/T_{0} \right) - 1 \right) }{ T_{0} - T_{f} } \right) t, & t < T_{0}  \\
                \left( \frac{s_{0} - 1}{T_{0}-T_{f}} \right) t + \left( \frac{1 - s_{0} \left( T_{f}/T_{0} \right) }{1 - \left( T_{f}/T_{0} \right) } \right), & t > T_{0} 
            \end{cases}
\end{align*}

\quad In the experiments on C17, $ T_{0} $ was set to $ 20 $ units and $ s_{0} $ was set to $ 3/4 $. We used the \texttt{DifferentialEquations.jl}~\cite{rackaukas2017differential} package in Julia~\cite{bezanson2017julia} to simulate the Schr{\"o}dinger equation for the time interval $ [ 0, T ] $ in this and the succeeding Section. Results in Fig.~\ref{fig:lin_v_param_v_best} suggests that while this parameterized function is not an optimal strategy (a strategy that optimally exploits the adiabatic condition is represented in the plot by opt\_adia), it has a clear advantage over the linear function and is therefore used in the next Section on benchmarks for the C26 circuit. 

\subsection{Generalized C26 Random Circuits: Effects of Degeneracy and Multi-Fault Solution Spaces}\label{subsec:26}
\begin{center}
	\begin{figure}
		\includegraphics[width=0.500\textwidth]{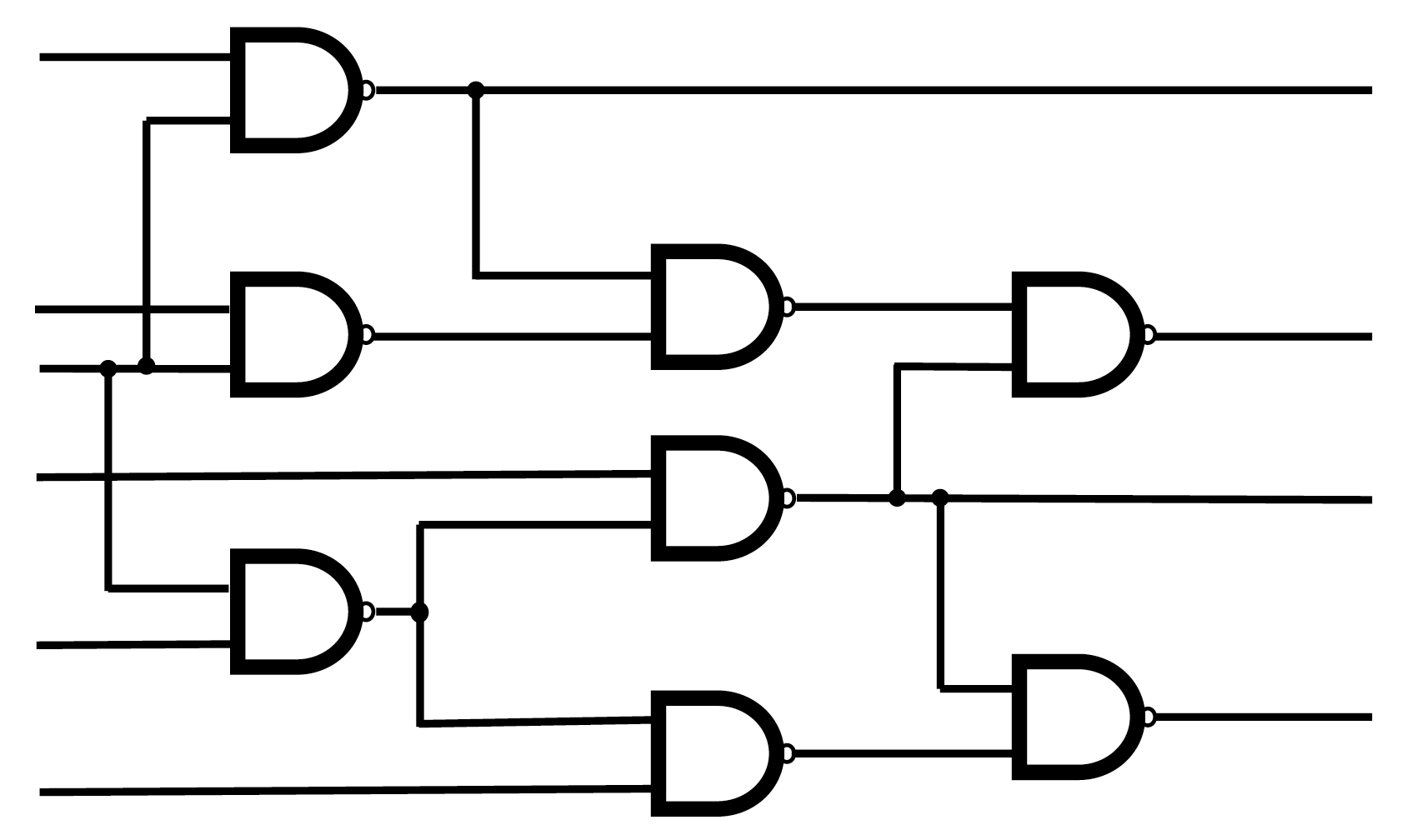}
		\caption{A schematic of the C26 circuit. For our C26 benchmarks, we replace each NAND gate with a randomly selected two input logic gate as well as randomly select inputs to the circuit. Each has 8 2-input logic gates, 6 FAN gates, 26 wires, 6 inputs, and 4 outputs.}\label{fig:c26}
	\end{figure}
\end{center}
\vspace{-2.5em}

\quad The number of outputs of a circuit is an upper bound on the number of faults of the minimum fault diagnosis. Since the C17 has only two outputs, if both of these are faulty, the MFD can have either 1 or 2 faults; the latter case is trivial (just assign the faults to the output bits), and the non-trivial case of a single fault is rather simple. In order to explore instances that offer higher complexity, we need to consider circuits with a larger number of outputs.  In this section, we consider a modification of C17 to a 26 wire, 6 input, 4 output circuit that we call C26 (see Fig.~\ref{fig:c26} for a schematic). 

\m 

\quad While the circuits studied in Sec.~\ref{subsec:c17} required 34 qubits to represent explicitly, the evolution of the wavefunction is restricted to a subspace of dimension $ 2^{15} $. In this section we consider a larger class of circuits that require 52 qubits to represent explicitly, but whose wavefunction evolution is restricted to a subspace of size $ 2^{22} $, which is still simulatable for sparse Hamiltonians.

\m 
\begin{center}
	\begin{figure}
		\begin{tabular}{ c c }
		\includegraphics[width=0.4600\textwidth]{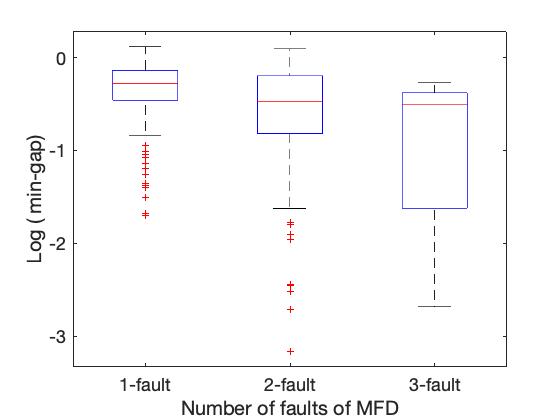} & \includegraphics[width=0.550\textwidth]{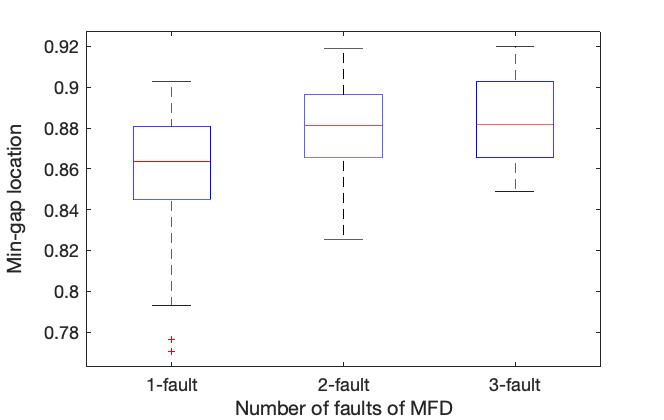} \\
		(1) & (2)
		\end{tabular}
		\caption{(1) shows a box and whisker plot of the logarithm of the minimum gap as a function of the number of faults in the MFD,  while (2) shows a box and whisker plot of the location of the minimum gap versus the MFD number of faults, for  randomly generated instances of C26 with non-degenerate ground states. }\label{fig:c26gapinfo}
	\end{figure}
\end{center}
\vspace{-2.5em}

\quad We generated random instances of C26 following the same procedure described in Sec.~\ref{subsec:c17}, and we selected 100 non-degenerate instances for each value (1, 2 and 3) of the number of faults in the MFD. We selected non-degenerate instances so that we could identify the location and size of the minimum gap (degenerate instances have a vanishing gap as $s \rightarrow 1$), and also because these are in some sense the hardest instances. In Fig.~\ref{fig:c26gapinfo}.1 we see that, even though the distribution of the minimum gap seems to spread towards smaller values as we increase the number of faults in the MFD, the median shows only a rather mild dependence. This means that an annealing schedule designed to solve, say, all non-outlier 1-fault MFD instances, will also solve about 75\% and just above 50\% of 2-fault and 3-fault MFD instances, respectively.   In Fig.~\ref{fig:c26gapinfo}.2, we also see that the location of the gap (similar to the C17 case) occurs in a well defined range for most instances and this appears stable  with regard to the number of faults of the MFD.

\m 

\quad However, as we consider larger circuits, the degeneracy of the MFD subspace will also potentially increase. That could be beneficial for a quantum annealing approach, since jumps from the ground state to excited states may still result in optimal solutions at the end of the anneal. Of course, we cannot know a priori whether a given instance is degenerate or non-degenerate, and what is the number of faults of its MFD. In order to gain some more insight into the performance of our approach on generic CFD instances, we simulated the closed system evolution and computed the success probability for a new set of randomly generated instances of CFD for C26, without implementing any type of post-selection. The only feature we exploited was the generic form of the spectrum (as seen in Fig.~\ref{fig:spectrum_stoq_n_nonstoq}.1), and implemented the modified annealing schedule  detailed in Sec.~\ref{subsec:c17}.

\m

\begin{center}
	\begin{figure}
		\begin{tabular}{c c}
\includegraphics[width=0.400\textwidth]{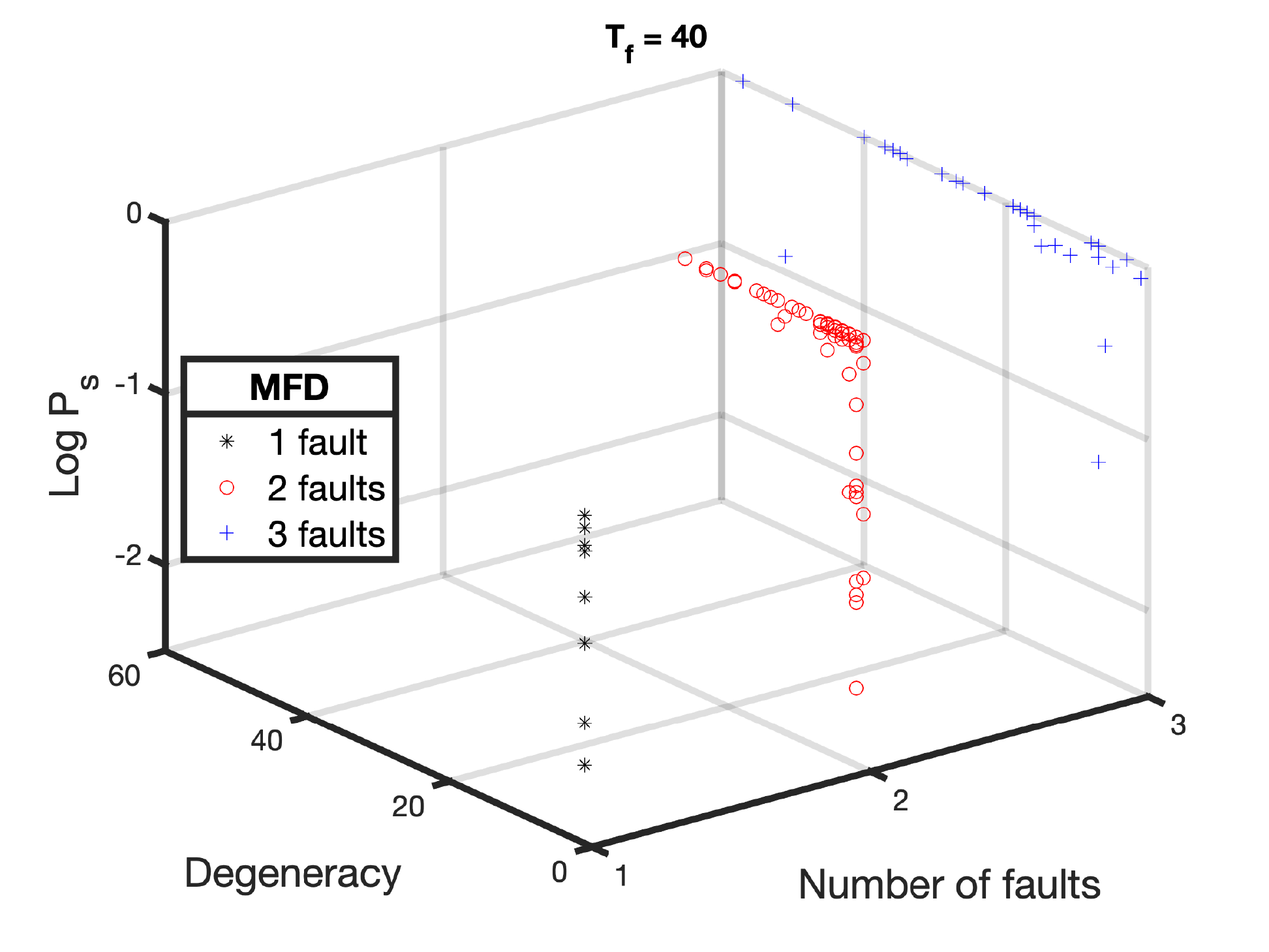} & \includegraphics[width=0.400\textwidth]{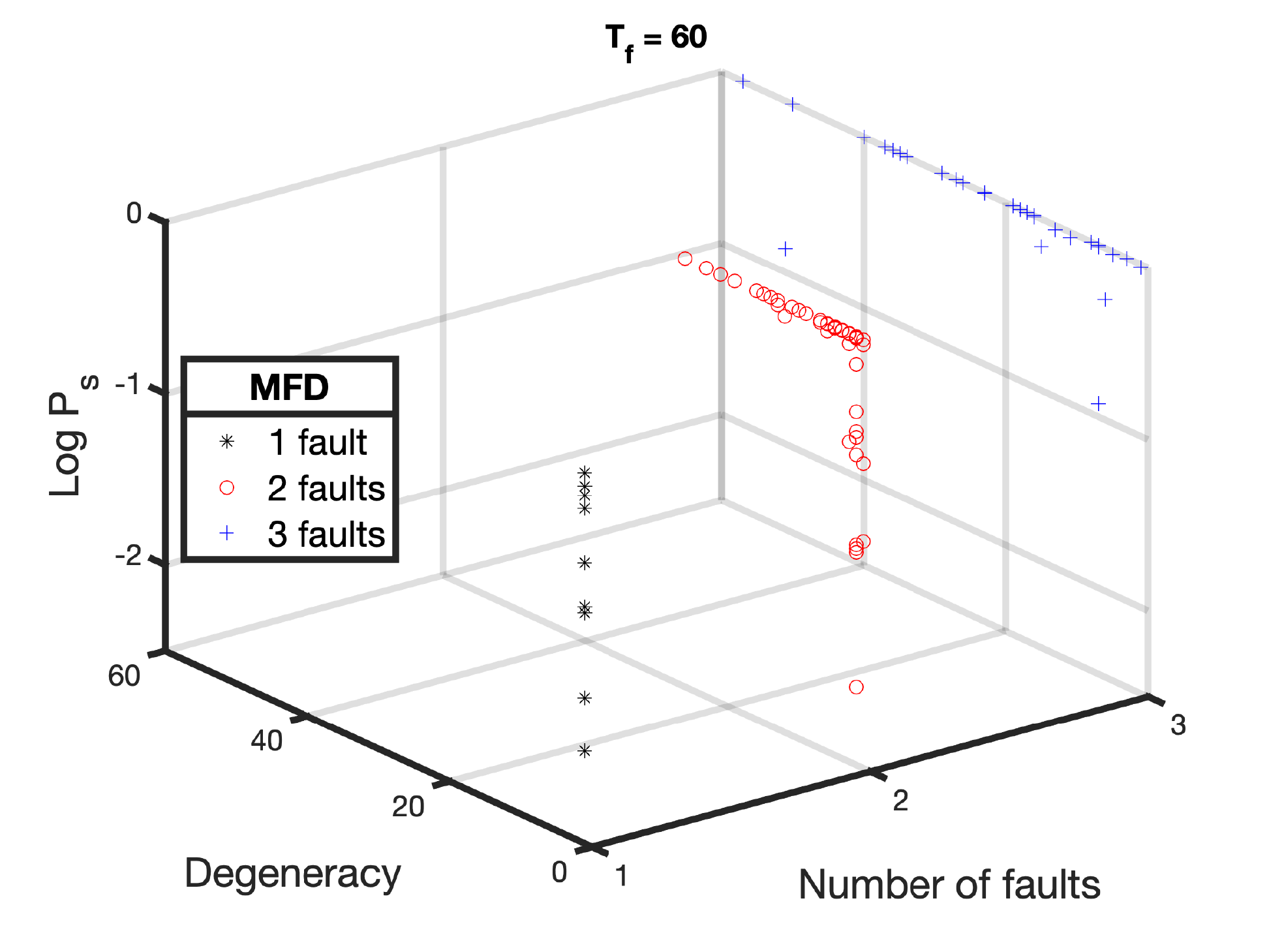} \\
		(1) & (2) \\
		\includegraphics[width=0.400\textwidth]{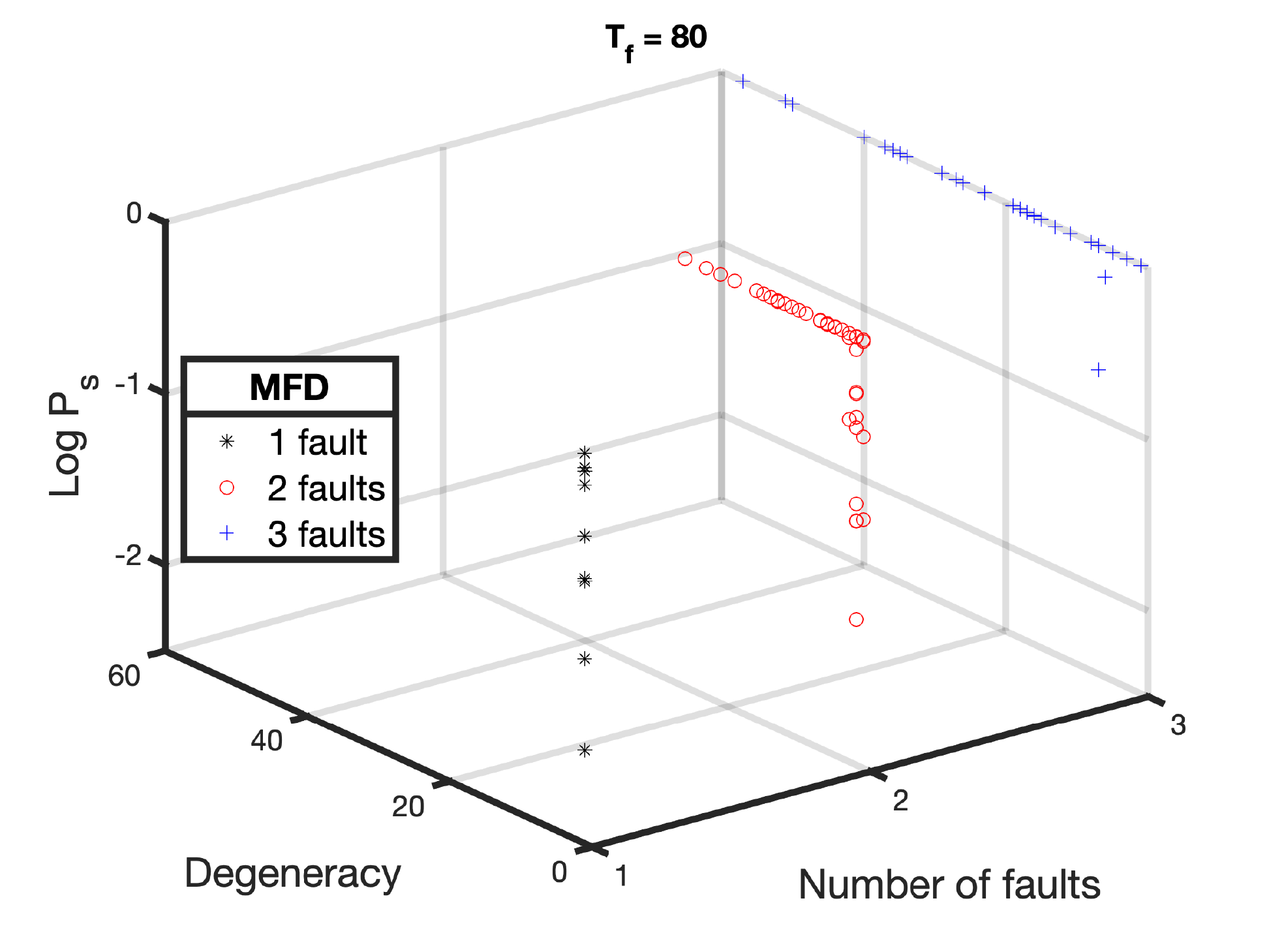} & \includegraphics[width=0.400\textwidth]{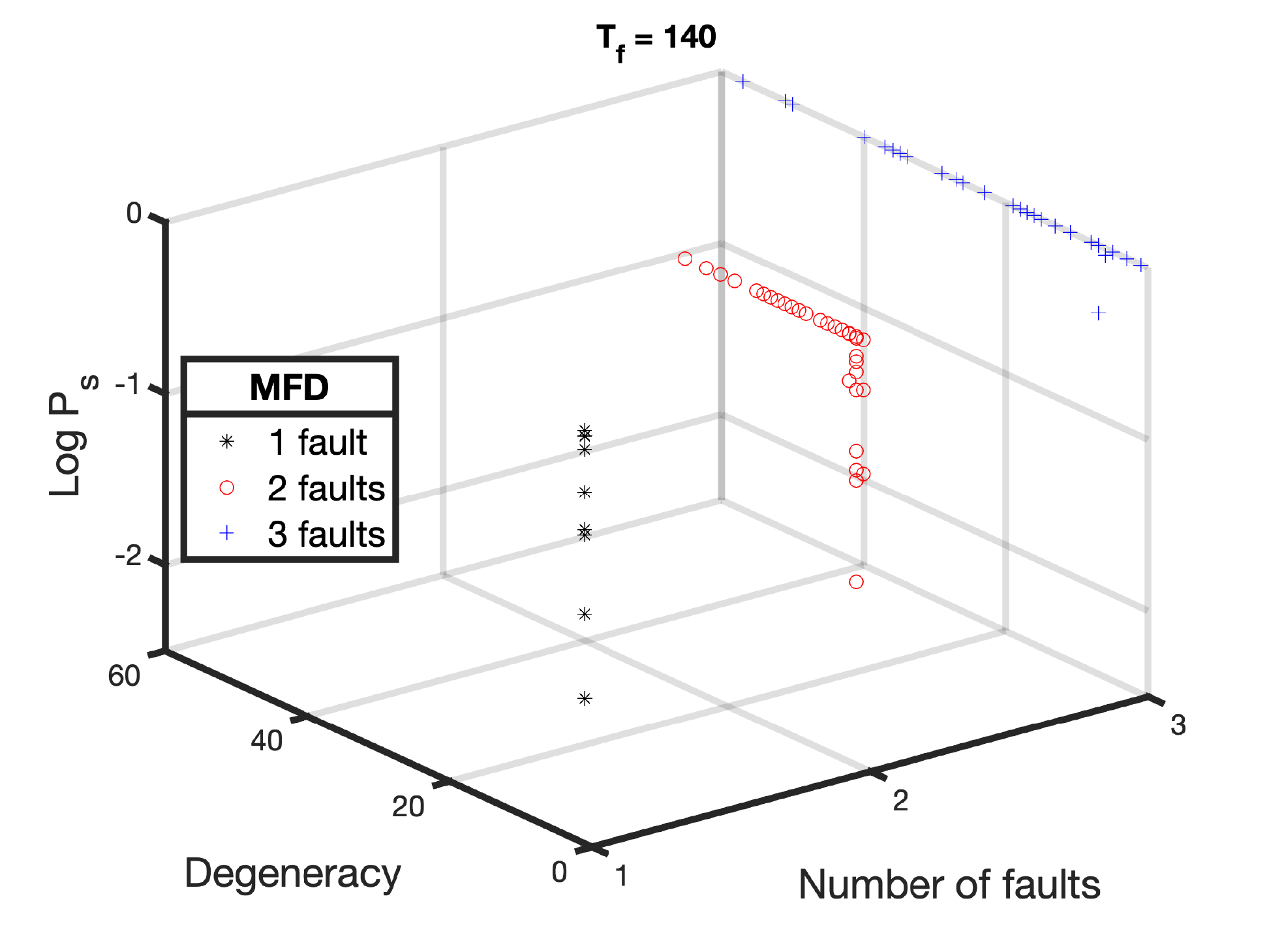} \\
		(3) & (4)
		\end{tabular}
		\caption{Logarithm of success probability as a function of instance degeneracy and number of faults in the MFD (for different values of the total annealing time $T_f$). }\label{fig:c26_degen_vs_succ}
	\end{figure}
\end{center}

\quad We simulated the annealing protocol for 100 randomly generated  instances (including those with degenerate ground spaces) for a total annealing time of 40, 60, 80, and 140 units. Note that (following our parameterized annealing schedule) 20, 40, 60, and 120 time units were consumed on the interval $ [3/4, 1] $, using a linear function,  and always 20 time units were used on the interval $ [ 0, 3/4 ] $ using a quadratic function.  Fig.~\ref{fig:c26_degen_vs_succ} shows the results of these experiments. First, we note that random instance seem  more likely to have 2-fault and 3-fault MFDs, and these in turn  tend to be more degenerate. Instances with a 1-fault MFD were always non-degenerate, and were the hardest instances to solve. In general, we can see that instances with low degeneracy are typically harder than those with high degeneracy. From a quantum annealing point of view we can understand this as results of the many jumps the system must go through to go from the ground state to the first state that is not part of the ground state manifold (i.e., the set of eigenstates that will end up converging into the ground state of the final Hamiltonian). High degeneracy seems to add extra protection against the effects of diabatic transitions, resulting in higher success probabilities.

\vspace{-2.5em}

\section{Conclusions}

\quad In this work, we introduced a novel approach to solving the CFD problem, where the quantum state throughout the anneal is constrained to the space of valid diagnoses. of size $ 2^{N-N_{O}} $ (rather than $ 2^{2N} $ for the total space). Previously studied QA approaches have relied on a transverse field driver Hamiltonians to solve CFD problems~\cite{bian2016mapping}\cite{perdomo2015quantum}\cite{perdomo2019readiness}. As noted in Ref.~\cite{bian2016mapping}, more novel driver Hamiltonians are one of the clearest routes to improving the performance for QA on such classes of problems and this work can be considered a step in this direction, relying specifically on the development and theory of CQA~\cite{hen2016quantum}\cite{hen2016driver}\cite{leipold2021constructing}.  Ordinary transverse field quantum annealing approaches typically require imposing penalty terms to suppress configurations that do not satisfy the problem constraints. Our approach exploits the structure of the problem to construct a set of special drivers (whose number is linear on the size of the circuit) that naturally implement transitions only between valid configurations. 

\m 

\quad We constructed a family of circuit instances generalizing the ISCAS C17 (17 wires, 5 inputs, 2 outputs) and analyzed the performance of the delineated protocol on these instances. By analyzing the spectral gap of a set of random instances, we were able to exploit its features (mainly the approximate location of the minimum gap) to construct an annealing schedule that evolved faster in regions of large spectral gap.  This information was then utilized to benchmark the performance of the approach on a family of larger circuits with 26 wires, 6 inputs, and 4 outputs (which we refer to as C26). We looked at the minimum gap for instances with non-degenerate solutions (i.e., those having unique solutions with the minimum number of faults), and found that the size of this gap had a mild dependence with the number of faults in the MFD. This is of interest, because large circuit ($\sim10^6$ wires) CFD problems typically found in industrial applications are currently solved using a brute force search approach, by looking for diagnosis with a fixed number of faults (typically 1 or 2). Since the complexity of this approach scales exponentially with the number of faults, it is currently not practical to search for diagnoses with higher number of faults. In contrast, the mild dependence of the minimum gap on the number of faults of the MFD suggests that our QA based approach could solve  some instances with more faults in their MFDs, using a fixed amount of computational resources, since QA is essentially characterized by the minimum gap. 

\m 

\quad In our QA approach based on special drivers, we chose the sign of each term in a way to make the overall Hamiltonian stoquastic. We considered one possible way of making it non-stoquastic by changing the sign of each driver term. Since non-stoquastic Hamiltonians could be more difficult to simulate, it has been speculated that they could provide more computational power. In our particular case, computing the spectral gap showed that  the minimum gap was typically smaller than for the stoquastic case, which would translate into a lower success probability for solving our CFD problem. However, it is important to point out that there are many ways of setting the phases of the driver terms in order to make the Hamiltonian non-stoquastic. A different choice (maybe informed by the circuit structure) could induce more interference effects between different diagnoses during the evolution, thereby possibly enhancing those with a lower number of faults. This is an intriguing possibility that we plan to pursue in future work.

\m

\quad In analyzing the performance of our approach, we focused on mainly two aspects we believe could prove beneficial when applied to larger scale problems. One was exploiting the generic form of the minimum gap as a function of the annealing parameter. We showed that it starts large, decreases almost monotonically and reaches a minimum in the last third of the anneal. These generic features allowed for a tailored schedule that moves faster at the beginning and slows down at the end where the minimum gap is typically located. We showed that this improves the probability of success for a given total anneal time, which will reduce the time needed to see a solution (i.e., the number of times the annealing should be repeated). 

\m

\quad The other aspect we studied was the dependence of the minimum gap on the number of faults in the MFD (minimum fault diagnosis). Our simulations showed that the dependence is rather mild, and furthermore, the increased degeneracy seen in instances with a higher number of faults in their MFD translated to a higher probability of success for instances with larger number of faults in their MFD. This is of interest because in practice (i.e., in actual industrial applications), large instances of CFD are tackled using exhaustive search for diagnoses with low number of faults (i.e., a few). Since the complexity of such search increases exponentially with the number of faults, this can become impractical for circuits with $10^6$, wires and diagnoses with more than 2 faults. Our analyses suggest that a QA approach could push this boundary further, resulting in significant practical (economical) impact. 

\m

\quad We would like to address the lack of head-to-head comparisons of our approach with classical algorithms. Our analysis involved  simulating the quantum evolution of systems that required considerable computational effort (C26 was represented using 22 qubits in the reduced subspace). However, these circuits are still small for classical methods: since the number of faults is upper bounded by the number of circuits outputs, our examples were still in the regime where exhaustive search would be extremely fast and classical algorithms more closely related to QA, such as simulated annealing, can be misleading to make comparisons with. Hence, we considered that a head-to-head comparison of time to solution for example, would not be very illuminating. Instead, as discussed in the previous two paragraphs, we decided to focus our analysis on features of the QA approach that we believe could provide an improvement when applied to larger circuits, where the combinatorial hardness of CFD becomes unavoidable for classical algorithms.

\m 

\quad Finally, we would like to note that our approach, even though designed with a quantum annealing framework in mind, can be translated to the circuit model of quantum computing, following the framework of the QAOA approach. For example, the construction of individual driver terms around gates can be translated into unitary mixing terms for the ansatz strategy based on the same underlying structure. Furthermore, since the structure of these driver terms is related to the circuit structure, it could be possible to gain some insight into better ways of choosing the mixing parameters, which could improve the performance of the QAOA approach. We consider this an intriguing direction for future research.

\section{Acknowledgements}

\quad FS thanks Sandeep Gupta for many useful conversations about the CFD problem. The research is based upon work (partially) supported by the Office of
the Director of National Intelligence (ODNI), Intelligence Advanced
Research Projects Activity (IARPA) and the Defense Advanced Research Projects Agency (DARPA), via the U.S. Army Research Office
contract W911NF-17-C-0050. The views and conclusions contained herein are
those of the authors and should not be interpreted as necessarily
representing the official policies or endorsements, either expressed or
implied, of the ODNI, IARPA, or the U.S. Government. The U.S. Government
is authorized to reproduce and distribute reprints for Governmental
purposes notwithstanding any copyright annotation thereon.

\m 

\quad The datasets used and/or analyzed during the current study are available from the corresponding author upon reasonable request.

\bibliographystyle{apsrev4-1}
\bibliography{main}

\end{document}